# *Ultrafast Reservoir Computing based on Nonlinear Nanomechanical Resonators at Ambient Conditions*

Enise Kartal[1,2,†], Yunus Selcuk[1,†], Batuhan E. Kaynak[1,2], M. Taha Yildiz[1], Cenk Yanik[3], M. Selim Hanay[1,2,*]

[1] *Department of Mechanical Engineering, Bilkent University, 06800 Ankara Turkey*

[2] *UNAM — Institute of Materials Science and Nanotechnology, Bilkent University, 06800 Ankara Turkey*

[3] *SUNUM, Sabancı University Nanotechnology Research and Application Center, 34956 Istanbul Turkey*

[†] *These authors contributed equally to this work.*

[*] *Corresponding author: selimhanay@bilkent.edu.tr*

**ABSTRACT**

Reservoir computing offers an energy-efficient alternative to deep neural networks (DNNs) by replacing complex hidden layers with a fixed nonlinear system and training only the final layer. This work investigates nanoelectromechanical system (NEMS) resonators for reservoir computing, utilizing inherent nonlinearities and the fading memory effect from NEMS's transient response. This approach transforms input data into a higher-dimensional space for effective classification. The smaller size and higher operating frequencies of the NEMS resonators enable faster processing rates than recent demonstrations with micromechanical systems, while their compact footprint and ability to operate under ambient conditions simplify integration into practical applications. Through an MNIST handwritten digit recognition test, this system achieved 90% accuracy with a 3.3-microsecond processing time per pixel, highlighting the potential for various applications that require efficient and fast information processing in resource-constrained environments.

**Introduction**

The development of increasingly complex algorithms and data processing tasks have pushed the boundaries of traditional computational models. While machine learning frameworks such as Recurrent Neural Networks (RNNs) offer impressive capabilities, they often require substantial computing power for training deep neural network architectures. The training becomes increasingly challenging in applications where miniaturized devices and low power consumption are essential, such as for applications in robotics and Internet of Things [1]. Resource-constrained edge computation devices operating within large networks necessitate efficient on-device processing and intelligent data analysis capabilities [2]. Here, limitations in processing power, memory, and battery life restrict the complexity of algorithms that can be deployed on these miniature devices.

Reservoir computing (RC) emerges as a potential solution to the aforementioned challenges [3-5]. Unlike traditional RNNs that require significant computational resources for training hidden layers, RC benefits from a high-dimensional reservoir of nonlinear elements to map input data to desired outputs [1]. This process makes RC computationally efficient, providing a framework desirable for resource-constrained edge devices, enabling on-device processing and intelligent data analysis [6] all without sacrificing precious device size or power consumption.

Over the past decade, reservoir computing frameworks have diversified significantly. Pioneering work relied on software implementations in conventional digital systems to define a nonlinear reservoir, such as Echo State Networks (ESNs) [3] and Liquid State Machines [4]. More recent advancements have focused on enhancing the performance of RC architecture and exploring unconventional physical and materials systems to serve as a rich reservoir [7-9]. A promising research direction explores the potential of nonlinear physical systems as computational reservoirs [10, 11]. In the mechanical domain, nonlinear Microelectromechanical Systems (MEMS) resonators with delay-coupled feedback were used to create virtual nodes in time and achieve reservoir computing in parity check and spoken word recognition tasks. Recently, the transient and nonlinear response of MEMS resonators were used for implementing reservoir computing without the need for a delay unit [12], and the ensuing device was applied to a variety of different classification tasks. These explorations pave the way for a unique synergy between reservoir computing and miniaturized mechanical systems (MEMS and NEMS), potentially leading to compact, low-power intelligent sensors with embedded reservoir functionalities.

The use of NEMS as the computational reservoir within RC presents a significant opportunity for developing miniaturized intelligent sensors. Unlike their larger MEMS counterparts, NEMS operates at the nanoscale, offering the potential for significantly reduced power consumption due to their inherent miniaturized nature. The mechanical resonances and easily accessible nonlinear regimes [13-17] in NEMS offer unique computational advantages as a reservoir, potentially leading to susceptible and efficient sensors capable of performing real-time data processing and analysis directly on the device.

This work investigates the application of a single nonlinear NEMS device as a computational reservoir within an RC framework (Figure 1). The input data is serialized and directly fed into the NEMS reservoir, leveraging its nonlinearity. In addition to nonlinearity, NEMS resonators provide the necessary fading memory effect owing to their transient response: information fed into the NEMS device at different times will have compounding effects due to the non-zero ring-down time of the resonator. These properties in principle facilitate the mapping of the input signal into a high-dimensional space where distinct classes become linearly separable. The input signal sampling time is maintained below or close to the ringdown time of the system to exploit the transient dynamics of the NEMS [12]. This configuration ensures that the reservoir response reflects a combined influence of past and current input values. The experiments were conducted under atmospheric pressure and required only standard electronics — features which enable a straightforward integration process for practical applications. Furthermore, using NEMS devices with high resonance frequencies facilitates operation on a significantly faster timescale than state-of-the-art mechanical devices, leading to a demonstrably quicker system response.

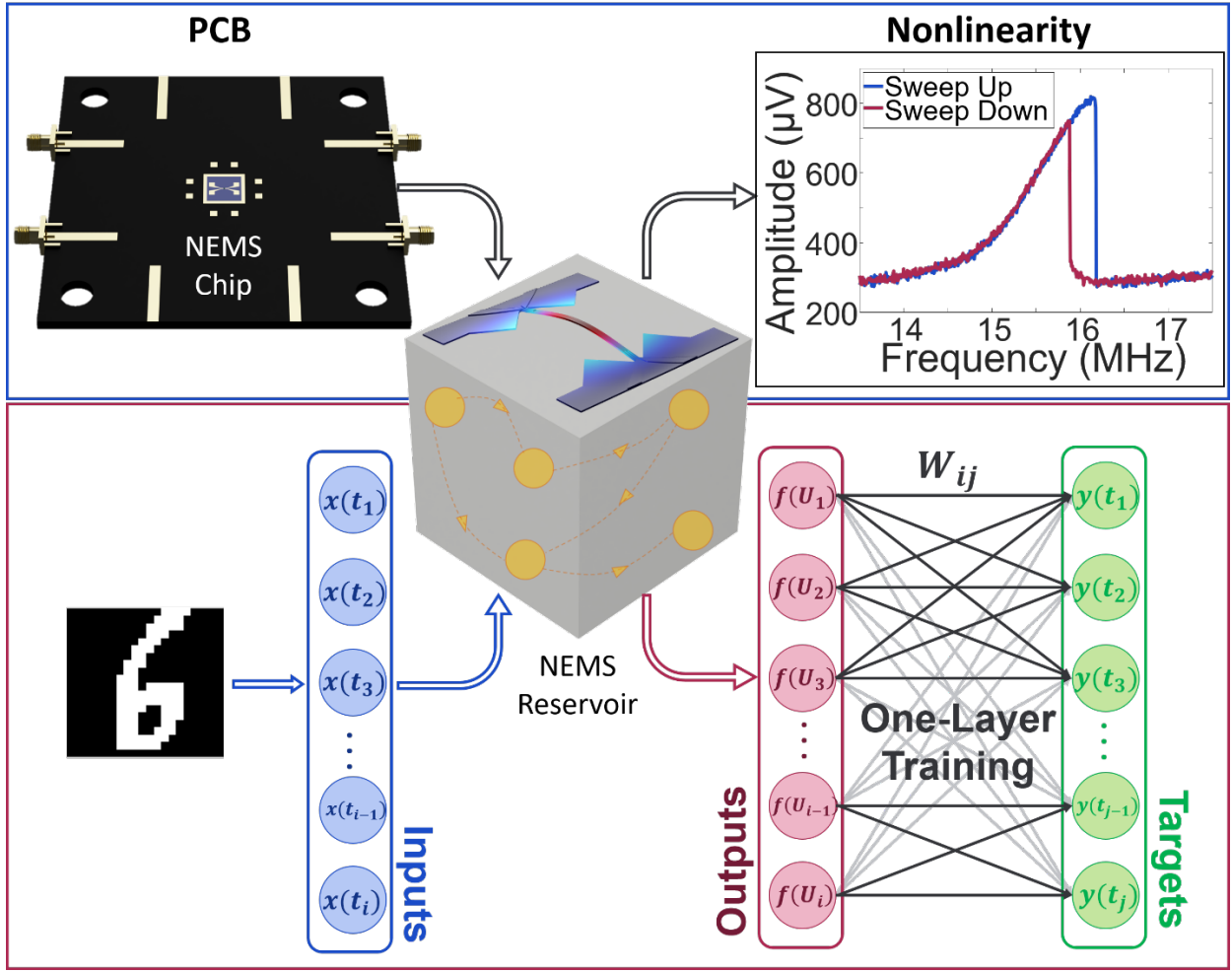

**Figure 1.** Reservoir computing architecture. Schematic of the nonlinear nanomechanical resonator-based reservoir computing system. The targets are the linear weighted sums of the output nodes of the reservoir. The nodes in the reservoir are the neural nodes created by the transient response of NEMS, which are connected temporally.

**Results/Discussion**

**Reservoir Computing System Based on a Nonlinear Nanomechanical Resonator**

Traditional time-delayed RC requires careful optimization of various parameters to generate a reservoir state with rich representational power, which ultimately determines the effectiveness of the system. In RC implementation, an input mask function serves as a critical element, which serializes the input signal and maximizes accessible dimensionality. In prior works, it was discussed that a self-masking procedure occurs within the reservoir. The masking procedure modulates the signal with arbitrarily fixed weights and feeds the response of the reservoir back to the system, thus creating an artificial fading memory. However, this process comes with various hyperparameters, and designing an optimal mask function can be challenging. Instead, the experimental setup of this work adopts a similar core concept developed in [12] where the masking procedure was removed. This simplification leads to exploring and focusing more on the intrinsic nonlinearities and fading memory of the MEMS in performing reservoir computations. Here we applied this technology to the smaller NEMS devices to benefit from their higher resonance frequencies and increase the processing speed.

The architecture consists of three main parts: an input layer, a reservoir, and an output layer. The input signals are preprocessed and vectorized before being fed into the reservoir to ensure compatibility with the NEMS reservoir, which operates on vectorized data. Subsequently, the normalized vector is fed as the modulation signal, where the carrier signal drives the NEMS resonator to the nonlinear region of its fundamental mode. This modulation serves to push the NEMS into nonlinear operational regime, a key factor for exploiting its information processing capabilities within the reservoir computing framework. Within the reservoir, the processing occurs, incorporating the input signal and the nonlinear system response. Finally, the reservoir states are sampled and used for training and testing procedures through a linear regression algorithm.

A crucial parameter in this process is the separation time, which dictates the frequency of updates on the input signal. In other words, separation timescale denotes the duration of each information unit (e.g. pixel value) presented to NEMS by modulating the drive signal near resonance. The selection of separation time is critical: choosing a separation time too close to the ringdown time ($\sim \frac{2Q}{f_n}$) diminishes the memory capacity of the output signal. Conversely, a minimal separation time ($< 0.1 \, t_{ring-down}$) hinders the ability of the reservoir to effectively map the input data into a higher-dimensional space, an essential step for subsequent classification tasks.

The proposed single-resonator reservoir computing (RC) experimental setup offers several advantages contributing to its practicality and efficiency. Firstly, the system operates entirely under atmospheric conditions, eliminating the need for complex vacuum systemss. The operation in ambient pressure reduces the Quality Factor: as a result, the ringdown time decreases, enabling for a fast system response. Importantly, NEMS resonator can still generate nonlinear response even when it possesses a low Quality Factor (~50) in ambient conditions. Operating at atmospheric conditions and with low power consumption provides an advantage for the potential of NEMS based RC systems in customer electronics, edge computation and internet of things.

Beyond its operational simplicity, the system benefits from the dynamical properties of the NEMS resonator. Due to its microsecond-scale ringdown time, training large datasets, can be completed within minutes including data pre-processing. Additionally, the compact device footprint facilitates seamless integration into diverse dimensional systems, highlighting its potential for broad applicability.

**Nonlinear Model of the NEMS Reservoir**

To implement a nonlinear RC system, we utilized a doubly-clamped silicon nitride beam (length = 10 μm, width = 400 nm, thickness = 100 nm) with a central platform (length = 2 μm, width = 3 μm, thickness = 100 nm) resonator as the core computational element (Figure 2). The natural frequency and quality factor ($Q$) of the NEMS significantly influence the performance of the RC system. A sufficient quality factor is preferred to ensure adequate memory capacity, as it directly influences the ringdown time $\left(T = \frac{2Q}{f_n}\right)$ of the resonator. Characterization of the nonlinear dynamic response of the beam is critical as it is the primary source of nonlinearity within the reservoir. Multiple frequency sweeps are conducted to identify suitable driving frequencies ($f_d$) and excitation amplitudes that produce the desired nonlinear state for achieving rich reservoir dynamics.

Figure 2a illustrates the nonlinear frequency response of the NEMS resonator. Figure 2b showcases the mode shape of the first resonance frequency used in the experiment. To guarantee a high signal-to-noise ratio and stable amplitude output, the NEMS was driven at a frequency ($f_d$ point) slightly lower than the frequency value corresponding to the front bifurcation point of the hysteresis loop, as shown in Figure 2a. Another reason of choosing a value lower than the front bifurcation point was to prevent switching between two stable branches during the experiments, since the predominant frequency drift in the experiments was towards lower values [16].

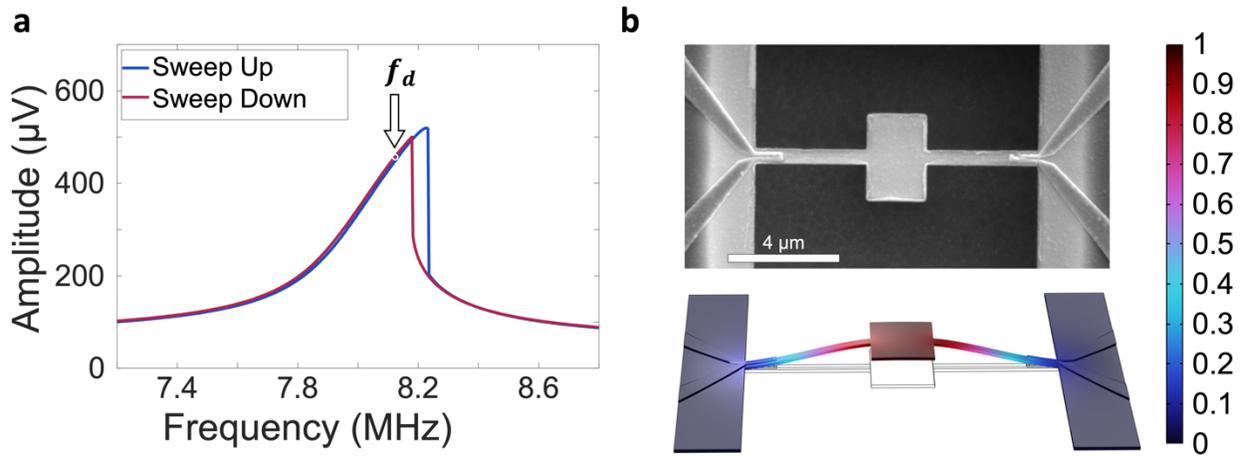

**Figure 2.** Nanoelectromechanical resonator and the first mode open-loop sweep. (a) Open-loop frequency sweep of the resonator at its first mode under atmospheric conditions. The frequency indicated by the arrow is the nonlinear driving frequency of the resonator, $8.12\ MHz$. (b) The scanning electron microscope (SEM) image of the device used in the experiments is shown in the top portion, and the COMSOL simulation for the fundamental mode used in the experiments. The colormap indicates the displacement amplitude.

**Experimental Setup**

The experimental setup, depicted in Figure 3, utilizes an arbitrary waveform generator (AWG) to create a modulated driving signal for the nanoelectromechanical resonator (NEMS) within the reservoir computing (RC) framework. This modulated signal incorporates the vectorized input data, containing pixel color information for each image in the dataset, and a sinusoidal wave with an amplitude and frequency corresponding chosen to set the NEMS in nonlinear driving regime. Essentially, the AWG performs amplitude modulation on the input data where the modulation amount is proportional to the brightness of the pixel being processed:

$$V_{signal} = \frac{(1 + m(t))}{2} V_d \cos(\omega_d t/2)$$

where $V_d$ is the initial nonlinear driving amplitude, and $m(t)$ contains the brightness information of each pixel normalized to be between 0 and 1, and the duration of each pixel information that is supplied to the NEMS device. The electronic signal driving the NEMS resonator ($\omega_d/2$) is provided at half of the mechanical driving frequency ($\omega_d/2$), since the thermoelastic actuation used here produces mechanical strain at double the excitation frequency [18, 19]. Similarly, the modulation term *m(t)* undergoes the square transformation by the electronic to mechanical transformation: we consider this drive-related nonlinearity (in addition to mechanical Duffing nonlinearity) as part of the total nonlinear response of the architecture.

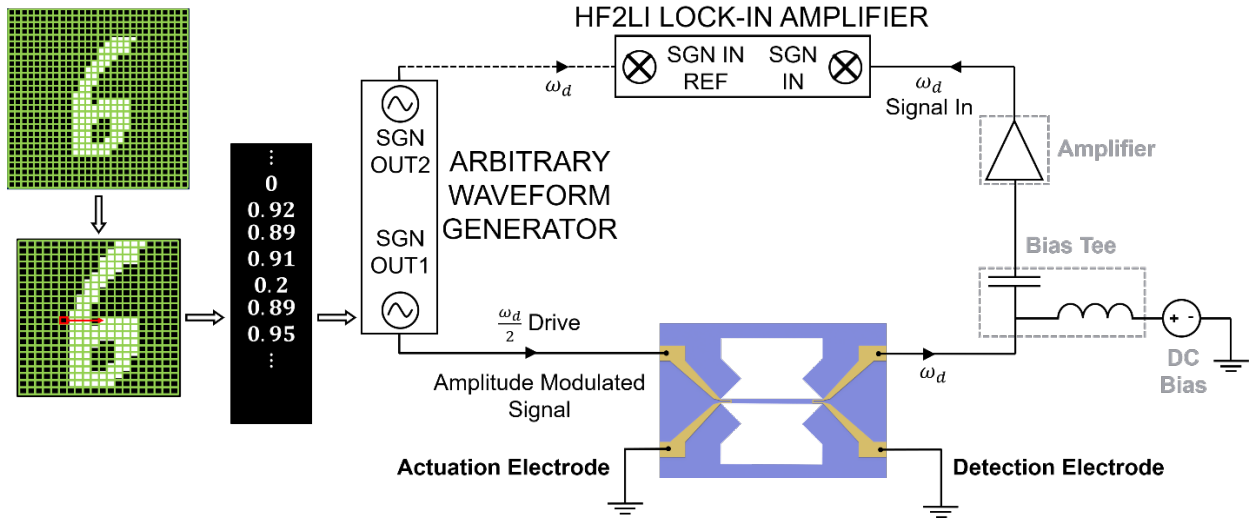

**Figure 3.** Schematic of the experimental setup and reservoir computing system. Cropped images with high information content are uploaded to the arbitrary waveform generator as vectors containing the grayscale values corresponding to the brightness of each pixel (normalized to unity for a fully white pixel). In the bottom middle of the circuit schematic, a render of the resonator can be seen. An AC signal (sinusoidal wave) modulated with the input vectors is fed to the actuation electrode of the resonator, and the reading is performed via the detection electrodes using a lock-in amplifier, which also receives an external reference signal (unmodified sinusoidal wave) from a phase synchronized second channel of the same AWG instrument.

After generating the signal modulated by the pixel values, the modulated signal drives the NESM resonator mechanically. The NEMS displacement is read out at the other end of the structure by the piezoresistive detection. The system employs a Lock-In Amplifier (LIA) for readout purposes. The LIA demodulates the output signal from the NEMS resonator using an external reference input provided by the second (and phase synchronized) output channel of the AWG at $\omega_d$: this reference input mirrors the second harmonic of the sinusoidal carrier (at $\omega_d/2$).. The demodulated output signal, containing the processed information, is then collected, and utilized for the training process.

This simple single-resonator RC experimental setup offers several advantages. It reduces overall complexity compared to most other RC systems while maintaining flexibility and improving information processing efficiency. The AWG's ability to seamlessly integrate the input data with the driving signal through amplitude modulation is another core benefit. Furthermore, the demodulation process facilitated by the LIA ensures the extraction of relevant information from the NEMS output, contributing to a streamlined and efficient information processing pipeline within the RC system.

**MNIST Handwritten Digit Recognition Task**

After obtaining the basic parameters of the NEMS resonator with the experimental system mentioned above, the classification performance of the system was evaluated using the MNIST handwritten digit dataset. The selection of the MNIST handwritten digit dataset for this research is motivated by the desire to evaluate the capability of the system for nonlinear information processing. MNIST, a widely recognized benchmark dataset for handwritten digit classification, comprises 28x28 pixel images offering a well-defined task for evaluating the nonlinear mapping properties of the NEMS reservoir (Figure 3).

The MNIST dataset images were cropped from their original size of 28x28 pixels to a smaller dimension of 22x20 = 440 pixels as a preprocessing step, removing potentially irrelevant blank peripheral regions. Preprocessing reduces redundant information within the input signals before feeding them into the reservoir (Figure 3) by cropping the images to their more informative parts. Subsequently, these cropped images are vectorized, resulting in a representation where each image is transformed into a 440-element vector. The form of supplied vector values of individual pixels, and the response of the nonlinear NEMS resonator are shown in Figure 4.

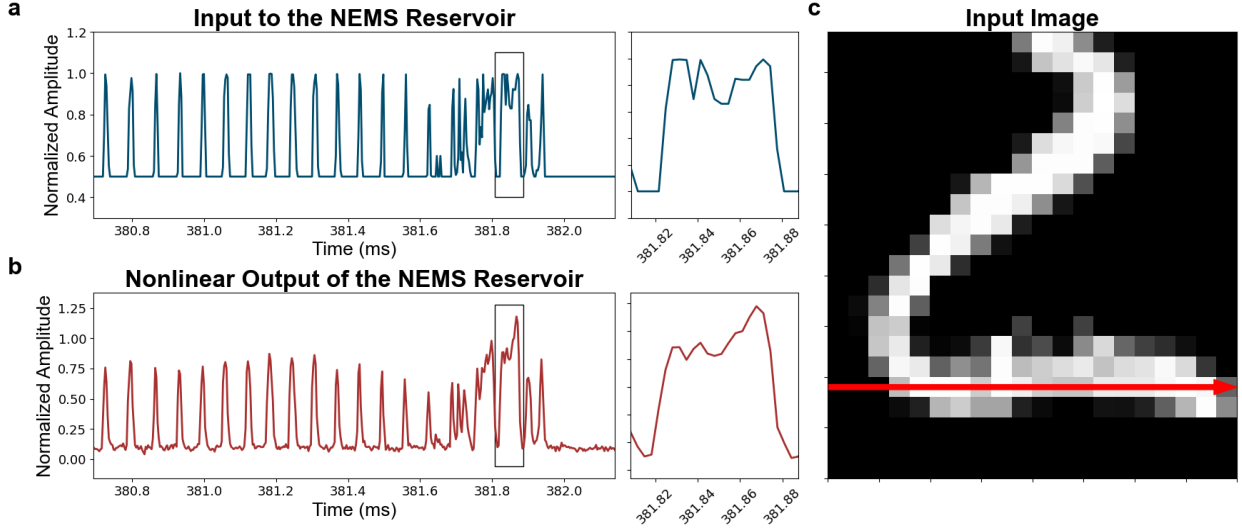

**Figure 4.** Comparison of signals supplied to and read out from the NEMS reservoir. **(a)** One of the images was provided to the NEMS reservoir as input. As the brightness of the pixel increases, so does the corresponding input drive level (y-axis). **(b)** The output signal after processed by the nonlinear NEMS resonator. **(c)** The specific image being processed by the NEMS. Red arow indicates the row being processed as highlighted in the insets.

The vectorization of the images facilitates compatibility with the experimental setup with the NEMS reservoir. As the serialized data (440 levels for each image) is fed into NEMS at the input port (Figure 4a), the NEMS output was collected in corresponding packages of 440 levels, forming the reservoir output features (Figure 4b). These 440 features are then connected to 10 output perceptrons, with identity activation function. The output layer is passed through a Softmax layer to obtain probability distribution of the predicted class of the input. Softmax provides good accuracy for problems with multiple classes and single labels, such as the digit recognition task. The training is done by optimizing the log-loss function with L2 regularization term to be small — *i.e.* until the loss function no longer changes by an arbitrarily set tolerance value (=0.0001 for our case) for 10 consecutive iterations. The loss function is expressed as:

$$L(p,q) = -\frac{\left(\sum_{j=0}^{n_{samples}} \sum_{i=0}^{n_{classes}} p(x_{ij}) \log\left(q(x_{ij})\right)\right) + L2 Regularization}{SampleSize}$$

where p is true probability distribution and q is predicted probability distribution and the L2 regularization term is expressed as:

$$L2 Regularization = \left( \alpha \sum_{n=0}^{n_{features}} \sum_{i=0}^{n_{samples}} w_{ni}^2 \right)$$

where α is the regularization strength constant and w is the weight between the NEMS output features and the output layers. The Softmax function is defined as follows:

$$Softmax(z_i) = \frac{exp(z_i)}{\sum_{j=0}^{n} exp(z_j)}$$

The weight optimization with respect to loss function is achieved by backpropagation algorithm with a learning rate of 0.001. To increase the efficiency, Adaptive Moment Estimation (Adam) solver was used. In total, we have utilized 24000 images from MNIST dataset, 2000 of which were randomly selected as test data. K-fold cross validation is applied for 60 folds and the average accuracy from the validation test can be seen in the Table 1.

A crucial factor influencing the reservoir dynamics of the NEMS is a specific nonlinear effect during information processing, which can be adjusted by varying the separation time. Separation time between two pixels can be defined with the parameter $\theta = n.T \ (n > 0)$ where T stands for the ringdown time of the NEMS resonator. The experiment was conducted at different separation times; however no significant effect was observed for the parameters tested in this study unlike the case with an earlier study [12]. In this work, the lock in time constant was kept constant at 0.77 microseconds for all different test conditions.

The ring-down time of the NEMS resonator (12.8 $\mu s$) sets the duration for which each image is processed. In this case, for the fastest processing of the data, each image is processed by NEMS within 3.2 microseconds and at 90% accuracy. A notable difference lies in the ring-down times observed in our study compared to existing literature [12]. A recent work of RC with no delayed feedback loop [10] reports significantly higher ring-down times, ranging from 16 to 35 milliseconds. Other works with low ring-down times close to the values reported in this work, such as the one presented in [20], depend on delayed feedback loops.

**Table 1:** Accuracy for MNIST digit recognition test for different separation times.

| $\theta = \dfrac{Separation\ Time}{Ringdown\ Time}$ | Separation time in microseconds | Average Training Accuracy | Average Test Accuracy |
|---|---|---|---|
| 0.25 | 3.3 | 93.0 % | 90.5 % |
| 0.5 | 6.6 | 93.2 % | 89.6 % |
| 1 | 13.2 | 92.6 % | 88.6 % |
| 2 | 26.4 | 94.3 % | 90.8 % |
| 4 | 52.8 | 94.5 % | 90.8 % |

These results demonstrate the effectiveness of the proposed RC system in exploiting the inherent nonlinearities of the NEMS reservoir for efficient information processing. The ability to adjust the nonlinear dynamic richness allows for tailored performance optimization based on specific classification tasks. The high classification accuracy achieved on the MNIST dataset demonstrates the proposed nonlinear RC system's efficient information processing for pattern recognition and its short-term memory potential for simple forecasting tasks. It suggests the feasibility of a simplified, single-resonator RC system operating under atmospheric conditions and improved computational speed compared to previous architectures.

**Conclusion**

The experimental analysis of the resonator's response and subsequent testing on the MNIST dataset demonstrate a classification test accuracy reaching above 90% for handwritten digits, suggesting the potential of this approach for facilitating hardware implementation of RC and paving the way for future applications leveraging NEMS technology. A comparison of the classification rates for different technologies and studies are summarized on Table 2.

This work presents a simplified, single-resonator reservoir computing (RC) system operating under atmospheric conditions that leverages the inherent nonlinearities of a NEMS resonator. This approach eliminates complex vacuum setups, enabling faster and more efficient experimentation compared to traditional RC systems with significantly longer ring-down times. Additionally, the NEMS resonator's ring-down decay time allows for experiments to be completed within minutes, and its compact size facilitates integration into various dimensional systems. These combined advantages, particularly the lower ring-down times, highlight the practicality and broad applicability of this novel RC architecture.

**Table 2:** A comparison of ring-down / relaxation time scales of recent reservoir computing work.

| Reference | Publication Year | Technology | Ring-down /Relaxation Time |
|---|---|---|---|
| **[12]** | 2021 | MEMS Resonator | ~ 16 milliseconds |
| **[20]** | 2021 | MEMS Resonator | few microseconds |
| **[21]** | 2021 | MEMS Resonator | ~ 330 microseconds |
| **[22]** | 2020 | MEMS Resonator | ~ 96 microseconds |
| **[23]** | 2022 | Photo-synaptic Memristor Array | 29 – 280 milliseconds |
| **[24]** | 2022 | 3D Memristor | ~ 600 microseconds |
| **[25]** | 2022 | 3D Dynamic Memristor Array | few microseconds |
| **[11]** | 2021 | 2D Memristor | ~ 100 milliseconds |
| **[10]** | 2021 | Dynamic Memristor | ~ 400 microseconds |
| **[26]** | 2020 | Memristor Network | ~ 100 milliseconds |
| **[27]** | 2019 | Magnetic Skyrmion Memristor | ~ 25 nanoseconds |
| **[28]** | 2022 | Optoelectronic Synapse on Van der Waals layer | 30 – 900 milliseconds |
| **[29]** | 2021 | Biocompatible Organic Electrochemical Network | ~ 100 milliseconds |
| **[30]** | 2023 | Ferroelectrics | 12 – 280 milliseconds |
| **[31]** | 2022 | Ferroelectric FET | ~ 100 nanoseconds |
| **[32]** | 2021 | Photonics | range of picoseconds to microseconds |
| **[33]** | 2021 | Spin-torque Nano-oscillator | ~ 200 nanoseconds |
| **[34]** | 2021 | Spintronics | 10 – 100 nanoseconds |
| **[35]** | 2020 | Spintronics | hundreds of nanoseconds |
| **[36]** | 2019 | Spin-torque Nano-oscillator | ~ 4 nanoseconds |
| **[37]** | 2021 | Stripe Magnetic Garnet | few nanoseconds |

**Methods/Experimental**

**Device Fabrication.** The doubly clamped beam resonator is fabricated on a low-pressure chemical vapor deposition (LPCVD) nitride layer on a silicon substrate (commercially obtained from University Wafer). The device dimensions are 10 μm length, 400 nm width, and 100 nm thickness. The u-shaped electrodes have an 80 nm width with an 80 nm gap. The fabrication starts with a two-step electron beam lithography (EBL) process: the first step defines the gold electrode pattern following gold deposition and lift-off, and the second step defines the copper etch mask pattern. Subsequently, copper is deposited and lifted off to serve as a dry etch mask for the nitride beam release. Then, an anisotropic inductively coupled plasma (ICP) etch is performed to etch the silicon nitride, and finally, an isotropic silicon etch is performed using ICP to achieve beam suspension. The copper mask is removed using a wet etchant, completing the device fabrication.

**Device Characterization.** The experimental setup for the single-resonator reservoir computing system utilizes a laboratory workstation, a Lock-In Amplifier (Zurich Instruments HF2LI), and an Arbitrary Waveform Generator (Keysight 33600A Series), and the fabricated nanoelectromechanical resonator device. The workstation operates LabOne software to control the LIA, while the AWG generates the amplitude-modulated input signal driving the resonator. Additionally, an amplifier, a bias-tee, and a DC power supply complete the circuit. Prior to conducting experiments with specific tasks, suitable driving parameters are identified. The LIA performs an open-loop frequency sweep to determine the optimal driving frequency and the effective quality factor of the system.

**Financial Interest Statement**: MSH is a cofounder in Sensonance Muhendislik, the other authors declare no competing interests.

AUTHOR INFORMATION

**Corresponding Author**

**Mehmet Selim Hanay** - Department of Mechanical Engineering, and UNAM — Institute of Materials Science and Nanotechnology, Bilkent University, 06800, Ankara, Turkey.

selimhanay@bilkent.edu.tr

**Present Addresses**

BEK and CY are now at California Institute of Technology, Pasadena, 91125, CA.

**Author Contributions**

BEK and MSH built the hardware architecture; BEK designed the devices; CY, EK and BEK fabricated the devices; EK, BEK and YS conducted the experiments; YS and EK optimized the experimental parameters and generated the plots; YS, MTY and EK conducted pre- and post-processing of the data; EK, BEK and MSH wrote the manuscript.

**Funding Sources**

This work was supported by the Scientific and Technological Research Council of Turkey (TÜBİTAK) via Grant EEEAG-122E564.

ACKNOWLEDGEMENT

We thank Mert Göksel and Alper Demir for useful discussions.